\documentstyle[12pt,fleqn]{article}
\input{epsf}

\newcommand{\be}{\begin{equation}}
\newcommand{\ee}{\end{equation}}

\newcommand{\noi}{\noindent}

\newcommand{\ga}{\alpha}
\newcommand{\gb}{\beta}

\newcommand{\gd}{\delta}

\newcommand{\gth}{\theta}

\newcommand{\gk}{\kappa}
\newcommand{\gl}{\lambda}

\newcommand{\gr}{\rho}

\newcommand{\gs}{\sigma}

\newcommand{\gff}{\varphi}

\newcommand{\gD}{\Delta}

\newcommand{\gW}{\Omega}

\newcommand{\ra}{\rightarrow}
\newcommand{\pa}{\partial}
\newcommand{\ov}{\overline}

\newcommand{\lan}{\langle}
\newcommand{\ran}{\rangle}

\title{Blue spectra and induced formation of primordial black holes}
\author{E. Kotok$^{1,2^*}$, P.Naselsky$^{1,3^*}$\\
\small
1) Theoretical Astrophysics Center, Juliane Maries Vej 30, 2100 Copenhagen, 
{\O} Denmark\\
\small
2)Keldysh Inst. of Appl. Mathem. Miusskaya sc.4, 125047 Moskow, Russia\\
\small
3) Rostov State University, Zorge 5, 344090 Rostov-Don, Russia\\
\small
-----------------------------------------------\\
\small
*) Permanent address}
\normalsize

\date{}
\begin{document} 
\maketitle

\begin{abstract}
We investigate the statistical properties of primordial black hole
(PBH) formation
in the very early Universe. We show that the high level of inhomogeneity of the
early Universe leads to the formation of the first generation PBHs.
This causes later the appearance of a dust-like phase of the cosmological
expansion. We discuss here a new mechanism for the second generation of
PBH formation during the
dust-like phase. This mechanism is based on the coagulation process.
We demonstrate that the blue power spectrum of initial adiabatic perturbations
after inflation leads to overproduction of  primordial black holes
with $10^9$g$\le M\le10^{15}$g if the power index is $n\ge1.2$.

\end{abstract}

\section {Introduction}

Blue primordial  power spectra of initial scalar perturbations have
spectral index $n>1$ and arise  naturally in the modern inflationary
scenario [1]. The observational upper limit on $n$ is derived by
normalizing the amplitude and spectral index $n$ to the quadrupole scale.
The COBE data [2] are compatible with a power spectrum of adiabatic
perturbations $P(k)\propto k^n$ with $n=1.2\pm 0.3$at 68\% confidential level (CL). At 95\% CL the $\frac{\Delta T}T$
observational limit on $n$ corresponds to $n<1.65$ [6].
It means that a
direct extrapolation of the COBE data to extremely small scales even with
the maximal possible value $n=1.5$ (one-sigma upper limit)
can give $\gd_{rms}$ great enough for the
formation of a large (remarkable) number of mini black holes [3,4].
Limits on their production [5] can be used to re-evaluate the restrictions
on the density perturbations spectrum including the spectral index $n$.
This problem was discussed by Carr, Gilbert and Lidsey [3], and recently
by Green and Liddle [4]. They found a significantly strong restriction on
the power spectrum index $n\le 1.3$, rather smaller than $n=1.5$
(one-sigma upper limit) from
COBE-data.

In our paper we demonstrate that due to the so called mechanism of
induced formation a blue spectrum of the initial
perturbations with $1.2\le n\le 1.3$ leads to the formation of a great
number of PBHs with $10^9$g$\le M \le 10^{15}$g

Firstly, following [3] we assume that the PBHs of extremely low mass
$M_{in}\ll 10^9$g
are formed directly in the early Universe from initial density
fluctuations with power spectrum $n\ge 1.2$ and COBE-normalized amplitude at
the moment $t_0\sim t_{Pl}(\frac{M}{M_{Pl}})$, where $t_{Pl}$ and $M_{Pl}$ are
the Planck parameters. We note that for blue spectra with $n> 1.2$ the
variance of fluctuations $\gs_{rms}^2$ is great enough for the formation
of a large number of mini black holes, which can produce a dust-like
phase of the expansion of the Universe at times $t\gg t_0$.

Secondly, we are interested in the situation where PBHs with
$10^9\le M \le 10^{15}$g are formed from PBHs with $M_{in}\ll 10^9$g due to
hierarchical clustering of fluctuations in the dust-like background of the
primordial low mass black holes $(M=M_{in})$. We demonstrate that during
a PBH dust era the fraction of the matter in the Universe accumulated
into PBHs with
$10^9\le M \le 10^{15}$g  is great enough to produce
the remarkable abundance of  the massive PBH
in the indicated interval of masses. These black holes will evaporate
up to the present time and will be able to change the observational
restrictions [5].

Finally, we shall demonstrate that for a blue power spectrum of
initial adiabatic
perturbations there is a limit on the power index: $n\le 1.23$.

The plan of this paper is as follows. In Section 2 we briefly discuss the
conditions for PBH formation in the early Universe with the equation of
state $p=\gamma\gr$ and $\gamma=1/3$. In Section 3 we investigate the
possibility to have a dust-like phase of the expansion of the Universe
due to PBH creation with extremely low masses $M_{in}$.
In Section 4 we discuss the conditions for formation of PBHs by
direct collapse of the initial perturbations.
In Section 5 we investigate a more efficient mechanism  of PBH formation
during the dust-like phase of the expansion of the Universe which is based
on the
coagulation approach.
In Section 6 we summarize the main results of the investigation.

\section{Formation of the PBHs with the extremely low mass.}

In this section we investigate  primordial low-mass
black holes formation from  density perturbations directly at the epoch
of the expansion  of the Universe with the hard equation of state
$p=\gamma\gr$. We assume that just after the period of reheating,
which is a natural part
of the modern theory of inflation, the equation of state corresponds to
$\gamma=1/3$ and the spectrum of the initial perturbations is blue:
$P(k)\sim k^n$; $n\le 1.5$ according to COBE data
(one-sigma upper limit) [1]. Following
[6] we assume that initial density perturbations produced during the
inflation period are Gaussian and the probability distribution of the smoothed
density field $P(\gd(M))$ is given by
\be
P(\gd(M))d\gd(M)=\frac{1}{\sqrt{2\pi}\gs_{rms}(M)}\exp
\left[-\frac{\gd^2(M)}{2\gs_{rms}^2(M)}\right]d\gd(M).
\ee
Here $\gs(M)$ is the mass variance evaluated at the horizon crossing;
\be
\gs^2(M)=\frac{1}{2\pi}\int^{\infty}_o dk k^2P(k)e^{-k^2R^2(M)}
\ee
where $P(k)=A^2k^n$ is the power spectrum with  amplitude $A^2$ and
spectral index $n$; $R(M)$ is the filtering scale  which is linked with
the mass of fluctuation $M_H=\frac{4\pi}{3}\gr H^{-3}$ at the horizon scale
$R\sim M^{1/3}$ for $\gamma=0$ of the equation of state and $R\sim M^{1/2}$
for the radiation-dominated Universe. In the  $\gamma=1/3$ Universe the mass
fraction of black holes $\gb(M)$, which are formed directly by the collapse of
initial perturbations at the horizon scale
can be obtain from formula (1) using Carr-Hawking
criterium $\frac{1}{3}\le \gd(M\sim M_H)\le 1$ (see [7]):
\be
\gb(M)=\int^1_{1/3}d\gd P[\gd(M)]\simeq \gs_{rms}(M)
\exp\left[-\frac{1}{18\gs_{rms}^2(M)}\right].
\ee

Note, that there are  a number of  observational limits on a PBH
fraction in the very early Universe, covering various mass ranges of its
mass spectrum (3).
For the future investigation we  use $\gb(M)$-limits from
nucleosynthesis and PBH evaporation at present time [5]. As one can see
from (3),  mass fraction of PBH
$\gb(M)=\frac{\gr_{pbh,in}(M_{in})}{\gr_{\gamma}}$,
is very sensitive to the amplitude and spectral
index $n$ of the initial density perturbations.
Here $\gr_{pbh,in}(M_{in})$ is the mass density of black holes and
$\gr_{\gamma}$ is the density of the ultrarelativistic background at the
moment of PBH creation.

For example, if $\gs_{rms}(M)\propto 10^{-5}$ and $n=1$ for Harrison=Zeldovich
spectrum normalized to COBE-data, then a probability of PBH formation is
extremely low: $\gb\simeq 10^{-6\cdot10^8}\sim 0$, and primordial black holes
practically cannot exist in the whole mass interval.
However, the situation changes drastically for blue power spectra
$n>1$.  It is worth to note  , that for $\gamma=1/3$ Universe the variance
of fluctuations at $M$ scale has a form [4]:
\be
\gs_{rms}^2(M)=\gs_{rms}^2(\ov{M}_{eq})
\left(\frac{M}{\ov{M}_{eq}}\right)^{-\frac{n-1}{2}};\hspace{0.1cm}
M<\ov{M}_{eq},
\ee
where ${M}_{eq}$ is the standard horizon mass at the moment of
matter-radiation equality. For $M>\ov{M}_{eq}$ up to COBE mass scales,
the variance is
\be
\gs_{rms}^2(M)=\gs_{rms}^2(\ov{M}_{eq})
\left(\frac{M}{\ov{M}_{eq}}\right)^{-\frac{n-1}{3}};\hspace{0.1cm}
M>\ov{M}_{eq}.
\ee
For the future investigation we use Eq.(4) and Eq.(5)
 and the fit of the four-year COBE data for the amplitude of
spectra at the scale $K_0=a_0H_0$ from [8]:
$$
A=\sqrt{2}\ \pi (a_0H_0) ^{-\frac{n+3}2}\delta _H,
$$
$$
\delta _H=1.9\times 10^{-5}\exp [ 1.01( 1-n) ] ,
$$
where $H_0$ and $a_0$ are the Hubble constant and scale factor at the
present time. The distribution of function $\sigma _{rms}\left( M\right) $
for different $M$, is plotted on the power index -mass diagram (Fig.1).
Further we use the modified power index $\alpha =\frac{n-1}4$ and
normalization $\sigma _{rms}\left( M\right)$ from Eq.(4)-Eq.(5) on the COBE
mass
scale $M_{COBE}$, which corresponds to $K_0=a_0H_0$:
\be
\sigma _{rms}\left( M\right) {\large =\sigma _{rms}\left( M_{COBE}\right)
\left( \frac{M_{COBE}}{\overline{M}_{eq}}\right) ^{\frac 23\alpha }\left(
\frac M{\overline{M}_{eq}}\right) ^{-\alpha },}
\ee
Let us introduce the new function $\alpha \left( M,\overline{\sigma }%
_{rms}\right) $:
\be
\alpha \left( M,\overline{\sigma }_{rms}\right) {\large =}\frac{\lg \left[
\frac{\overline{\sigma }_{rms}}{\overline{\sigma }_{rms}\left(
M_{COBE}\right) }\right] }{\lg \left( \frac{M_{COBE}}M\right) -\frac 13\lg
\left( \frac{M_{COBE}}{\overline{M}_{eq}}\right) }{\large ,}
\ee
where $\overline{\sigma }_{rms}=const$ - is the parameter.

\noindent
Note, that for any kind of blue power spectrum the power index $\alpha
=\alpha _{*}=const$. Let us consider the equation $\alpha \left(
M,\overline{%
\sigma }_{rms}\right) =\alpha _{*}$. For different values of the parameter
$\ov{\sigma }_{rms}$ this equation gives us the function
$M=M_{*}(\ov{\sigma }_{rms})$, which corresponds to the
$\sigma_{rms}(M)$ from Eq.(6)  at $\alpha =\alpha _{*}$.
Using this method, we would like to focus attention on the thin dashed,
dotted and solid lines of Fig.1, which are concentrated at the range of
$\alpha<0.09$ (see Fig.1). These lines correspond to the series of
function $\alpha(M,\ov{\sigma }_{rms})$ from Eq.(7) with the different
values of $\ov{\sigma }_{rms}$. The lowest dashed line corresponds to
$\ov{\sigma }_{rms}=0.020$, the next line (dotted) corresponds to
$\ov{\sigma }_{rms}=0.025$.

\begin{figure}[h]
\vspace{0.3cm}\hspace{-1cm}\epsfxsize=9cm
\epsfbox{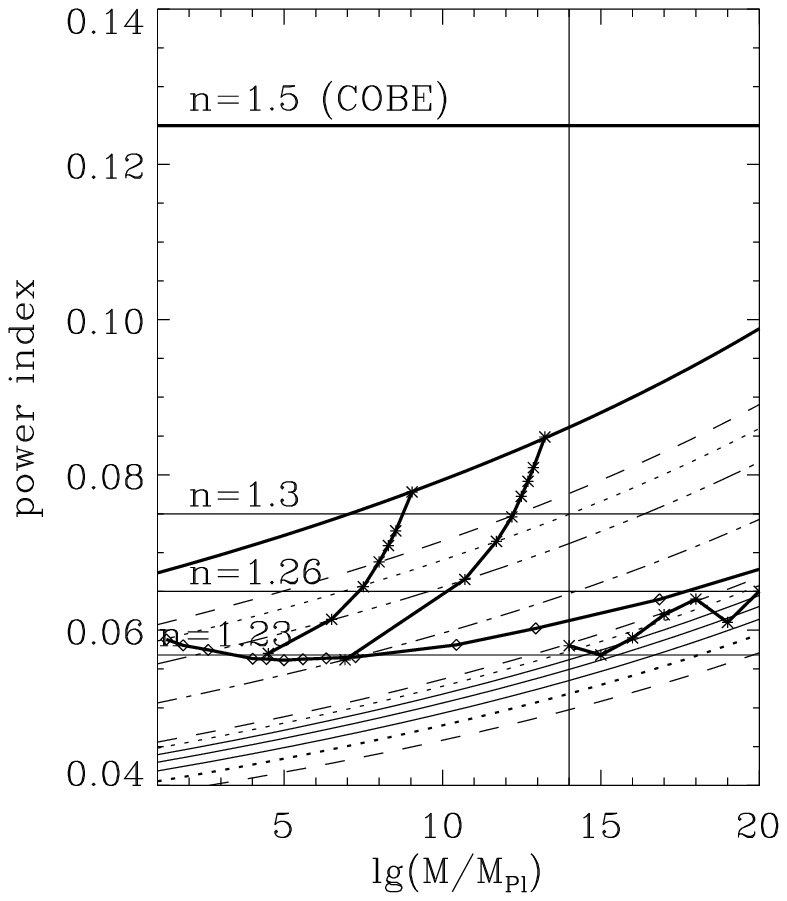}
    \caption{The diagram: the power spectrum index - mass.
 The tilted dashed, dotted, thin solid and so on lines
correspond to the different values of the parameter $\ov{\sigma}_{rms}$
of the function $\ga(M,\ov{\sigma}_{rms})$, see formula (7).
    For seven of these lines in the bottom part of the diagram,
from the bottom to the top the amplitude
of perturbations $\ov{\gs}_{rms}$ changes from $2\cdot 10^{-2}$ to $5\cdot
10^{-2}$
with the step $\gD=5\cdot 10^{-3}$.
The next series of the tilted lines (in the middle part of the diagram)
corresponds
to the levels of amplitude $\ov{\gs}_{rms}=0.1\div 0.4$ with $\gD=0.1$.
The last solid tilted line corresponds to the amplitude $\ov{\gs}_{rms}=1$.
Four horizontal lines from the top to the bottom correspond to $n=1.5$
(COBE-restrictions on the spectrum power index); $n=1.3; 1.26; 1.23$.
The solid line with asterisks inside the range $14\le\lg M/M_{Pl}\le 20$
corresponds to the observational restrictions [5].
The solid line with diamonds corresponds to the condition
$\tau_{col}=\tau_{ev}$ (see Section 5).
Two solid lines with asterisks in the range $5\le\lg M/M_{Pl}\le 14$
correspond to the condition $M_{BH}<10^9$g and $M_{BH}<10^{15}$g
correspondingly (see text). The vertical line $M=10^{14}M_{Pl}$ is the
lowest limit of mass for which there exist observational restrictions,
based on the Hawking's process. See the text for other details.
}
\end{figure}
\noi
Three next solid lines represent
$\ov{\sigma}_{rms}=0.030;0.035$ and $0.040$ correspondingly. Two next
lines (short dashed and long dashed) correspond to
$\ov{\sigma }_{rms}=0.045$ and $\ov{\sigma }_{rms}=0.050$.
As in can be seen from Eq. (3) and Eq.(7) for
this kind of function $\alpha(M,\ov{\sigma }_{rms})$ the
value of parameter $\ov{\sigma }_{rms}$ can be normalized on the
fraction $\beta $ of PBH (see Eq.(3)). Using this normalization, we
plotted in Fig.1 the thick solid line at the interval
$14\leq \lg M/M_{PL}\leq 20$
which corresponds to the well known limits of PBH's probabilities [5].

There is no observational restriction at $M<10^9$g, and we can use only one
limit $\gW_{pbh}<1$, where $\gW_{pbh}$ is the ratio of the black hole and
critical densities at present time, for $M\gg 10^{15}$g .

Note, that the increasing of the variance of fluctuations at $M\ra M_{Pl}$
is the natural behavior of the blue power spectrum. There is not such a
behavior
for Harrison-Zeldovich and tilted ($n<1$) power spectra.
The growth of $\gs_{rms}$ at small scales
of perturbations is a cause of the producing  low-mass PBHs ($M\ll 10^9$g but
$M\gg M_{Pl}$) which evaporate before the period of the cosmological
nucleosynthesis. However, these low-mass PBHs can produce the dust-like phase
of the expansion of the Universe and change the condition of formation
the next generation of PBHs with $10^9$g$\le M\le 10^{15}$g. Thus,
the high density of the low-mass PBHs (we use an index ``$in$'' for
this first
generation of PBHs) can induce the overproduction of the PBHs with
$10^9$g$\le M\le 10^{15}$g (the second generation) at the dust-like phase
 of the expansion.

In the next section we will discuss this possibility in some detail.

\section{Dust-like phase of formation of the PBHs.}

Let us describe the condition  which produces the PBH dust-like phase of
the expansion of the Universe. We denote by $t_{in}$ the time of formation
of black holes of the first generation in the radiation dominated Universe.

Let  $\beta(M_{in})$ be a fraction of matter to the Universe collapsed
into the first generation PBHs with  mass $M_{pbh}=M_{in}$.
It is well known [4, 6] that the
period of formation, $t_{in}$, and the mass of PBH $M=M_{in}$ are linked by
the equation:
\be
M_{in}=\gamma^{3/2}M_H=
\frac{\gamma^{3/2}}{g^{1/2}_f}\left(\frac{t_{in}}{t_{Pl}}\right)M_{Pl},
\ee
where $g_f$ is the effective number of massless degrees of freedom at
 $t=t_{in}$.
After primordial black holes are formed, the mass density of its first
generation $\gr_{pbh,in}$ begins to decrease  as
$\gr_{pbh,in}\propto a^{-3}$, where $a$ is a scale factor. The mass
density of ultrarelativistic matter decreases more rapidly:
$\gr_{\gamma}\propto a^{-4}$. In this case at
\be
t>t_{eq}, \hspace{1cm}t_{eq}\simeq t_{in}(M_{in})\gb^{-2}(M_{in})
\ee
the cosmological expansion is determined by PBHs with
$M=M_{in}$. If the time of evaporation of these black holes
\be
\tau_{ev}=
\frac{1.2\cdot10^4}{\gk}\left(\frac{M_{in}}{M_{Pl}}\right)^3\cdot t_{Pl},
\ee
is much greater than the time $t_{eq}$,
then the duration of PBH dust-like phase is
$\gD t=\tau_{ev}(M_{in})\gg t_{eq}(M_{in})$,
where $\gk$ is the parameter depending on the number of particles species
which can be emitted (see [4])).

It is worth to note that at the moment $t=t_{in}$ the space distribution of
PBHs with $M=M_{in}$ is modulated by  the long wave perturbations in the
ultrarelativistic background. This spectrum of adiabatic
fluctuations is preserved in the PBH's distribution up to the dust-like phase.

At $t>t_{eq}$ density fluctuations in the non-relativistic ``gas'' of
PBH growth with time and the variance of perturbations in the PBH's
distribution at $t\gg t_{eq}$ is:
\be
\gs_{rms}(M,t)=\gs_{rms}(M_{eq})T(M,M_{eq})\left(\frac{t}{t_{eq}}\right)^{2/3}
\ee
where $M$ is the mass of perturbations, and
\be
T(M,M_{eq})=\left\{\begin{array}{lll}
\left(\frac{M}{M_{eq}}\right)^{-\frac{2\ga}{3}} & {\normalsize{at}} & M\le
M_{eq}\\
\left(\frac{M}{M_{eq}}\right)^{-\frac{2(1+\ga)}{3}} & {\normalsize{at}} & M >
M_{eq}\\
\end{array}
\right.
\ee
is the transfer function of perturbations, and $M_{eq}$ is the mass of PBHs in
the volume $V\sim t^3_{eq}$:
\be
M_{eq}\simeq \gb^{-2}(M_{in})M_{in}
\ee
 As it is seen from (11) - (13), the spectrum $\gs_{rms}(M,t)$ is practically
flat ($\ga\ll 1$) when $M\le M_{eq}$ and there is a turn over point
at $M>M_{eq}$.
This  spectrum arises only under conditions $M_{in}\ll M_{eq}$ and
$\beta(M_{in})\ll 1$, when the hydrodynamic approach is correct. Actually
(see Fig.1) for the blue power spectrum with $n=1.23$ the level of
primordial fluctuations at small mass scales is comparatively high.
For example, if $M_{in}\approx 1$g, then $\gs_{rms}(M_{in}=1$g$)\approx 0.1$;
$\gb(M_{in}=1$g$)\approx 4.4\cdot10^{-4}$ and
$M_{eq}\approx5\cdot10^7M_{in}$.
In this case the spectrum of perturbations in the black holes background is
similar to well known spectral distribution
of fluctuations in the standard Cold Dark Matter (CDM) Universe model.

This analogy has a deep reason. In the standard CDM theory of the formation
of the large scale structure of the Universe the typical scale of the
mass density
perturbations has a turn over point at the scale of the equality,
which is close to
the typical scale of clusters of galaxies ($M_{eq}^{st}$). At the scale
$M<M_{eq}^{st}$ the dynamics of clusterisation of perturbations
is very complicated because of nonlinear transformation of
spectra which was investigated by Doroshkevich et al. [9].

According to [9], the hierarchical clusterisation of non-linear configuration
takes place at all scales of perturbations.  During this process
the typical configurations of low mass (mini pancakes) are clustering
step by step into  massive pancakes up to the typical scale
$M\approx M_{eq}^{st}$.
In the standard CDM Universe clusters of galaxies are formed during a very
short time corresponding to a redshift $z$, $0\le z\le1$.
It means that nonlinear transformation of the spectrum of perturbations has
a cutoff  caused by the limitation of the age of the Universe.

For the PBH's dust-like phase at $\tau_{ev}\gg t_{in}$ the situation changes
drastically. The duration of this phase covers not only the period of the
formation of structures with $M\approx M_{eq}$, but also the period of the
formation of structures with  $M=M_{max}\gg M_{eq}$. In this case
objects with $M\approx M_{eq}$ at the moment  of evaporation $\tau_{ev}$
are virialized and form the high massive objects with $M=M_{max}$
(see (11) - (12)):
\be
M_{max}\simeq M_{eq}\gs_{rms}^{-\frac{3}{2(1+\ga)}}(M_{in})
\left(\frac{\tau_{ev}(M_{in})}{t_{eq}}\right)^{\frac{1}{1+\ga}}.
\ee

In addition, there are two very important differences between the
rate of clusterisation during  PBH dust-like phase and the standard
CDM model of the structure formation. Firstly, if $\gb (M_{in})$ is not
especially small, then the scale of maximum of perturbations $M_{eq}$ contains
$N\le 10^2$ mini PBH with $M=M_{in}$, and for the description of the rate of
clusterisation one needs to use the kinetic approach. This situation takes
place, for example, if initial power index is $n=1.26$ and $M_{in}$=1g
(see Fig.1). As it is seen from Fig.1, $\gs_{rms}(M_{in}=1$g$)=0.3$;
$\gb(M_{in})\simeq 0.16$ and $N\simeq40$.

Secondly, in the context of the CDM scenario the background particles are
collisionless and ``interact'' each other and with the baryonic matter only
through gravitation.
However, PBHs with $M=M_{in}$ can collide especially inside the
quasi stationary stable gravitational
configurations (PBH's clusters - ``gas''), which are formed during the
dust-like phase.

These two last reasons are linked with a very important question -
can the first generation primordial black holes produce the next
generation of the PBH during the dust-like phase or they have to disappear
because of Hawking's evaporation without any observational consequences?

The answer this question depends very essentially on the probability
of PBH formation at the dust-like phase, which we describe in
the next section.

\section{The probability of the formation of the second generation of PBHs.
The spherical symmetry approximation.}

The problem of PBH formation during the dust-like phase was pointed out in the
pioneering articles by Polnarev and Khlopov [10,11]. According to [10] the
fraction of matter going into PBH is determined by the probability
that the overdense region is sufficiently spherical symmetric and
homogeneous. This region collapses into a black hole directly. The fraction
of PBH which is formed due to this mechanism is given by [10]:
\be
\gb(M_{bh})\simeq 2\cdot10^{-2}[\gs_{rms}(M,t_d)]^{13/2},
\ee
where $\gs_{rms}(M,t_d)$ is a variance of the density perturbations at
the moment $t_d$ when the dust-like phase starts.

Below we  briefly discuss the initial conditions leading to Eq.(15).
As we mentioned above, if the hydrodynamic approach is applicable,
we can describe the statistic properties of the Gaussian field of the
density perturbations according to Bardeen et al. [12].
First of all,
we can express the Gaussian point process entirely in terms of the
density perturbations field $\gd(\vec{r})$ and its derivatives.
In the vicinity of the point of a maximum of $\gd$ field $\vec{r}=\vec{r_*}$
we can use the Taylor series expansion:
\be
\gd(\vec{r})=\gd(\vec{r^*})+\frac{1}{2}\sum_{i,j}\gl_{i,j}(r_i-r^*_i),
(r_j-r^*_j)
\ee
where $\gl_{i,j}$ is the second derivative tensor of the field $\gd$
at $\vec{r}=\vec{r^*}$ (so called deformation tensor). We transform
the coordinate system along the principle axes of the deformation tensor
and determine  three its eigenvalues $\gl_1$, $\gl_2$ and $\gl_3$.
Doroshkevich [13] has shown that the joint probability density for
$\gl_1$, $\gl_2$ and $\gl_3$ has a form:
\be
P(\gl_1,\gl_2,\gl_3)=
\ee
\[
D(\gl_1-\gl_2)(\gl_1-\gl_3)(\gl_2-\gl_3) \exp
\left(-\frac{3\mu_1^2-7.5\mu_2}{\gs_{in}^2}\right),
\]
where $D=\frac{5^{5/2}\cdot27}{8\pi\gs_{in}^5}$, $\gs_{in}$ is the
variance of $\gd(\vec{r})$ at the moment $t_d$;
$\mu_1\equiv \sum_{i=1}^3\gl_i$ and
$\mu_2\equiv \gl_1\gl_2+\gl_2\gl_3+\gl_1\gl_3$.
Following [12] we define a three-axial ellipsoidal surface with semiaxes
\be
\xi_i=\left[\frac{2(\gd(0)-\gd(r))}{\gl_i}\right]^{1/2},
\ee
and characterize the asymmetry of $\gd(r)$ field at the vicinity of
the point of maxima ($r^*=0$) by the parameters:
\be
e=\frac{\gl_1-\gl_3}{2\mu_1}; \hspace{1cm} p=\frac{\mu_1-3\gl_2}{2\mu_1}.
\ee
As the result, the distribution of $\gd(r)$ field around a point of a
maximum in the spheric system of coordinates is [12]
\be
\gd(r)\simeq\gd(0)-x\gs_2\frac{r^2}{2}[1+D(e,p)]; \hspace{1cm} r\ra 0,
\ee
where
\be
\begin{array}{l}
D(e,p)=3e[1-\sin^2\gth(1+\sin^2\gff)]+p[1-3\sin^2\gth\cos^2\gff],\\
x=-\frac{\nabla^2\gd(r)}{\gs_2};\hspace{0.2cm}
\gs^2_n=\frac{1}{2\pi^2}\int dkk^{2(1+n)}P(k)\\
\end{array}
\ee
$P(k)$ is the power spectrum of the density perturbations.

The main idea which was used by Polnarev and Khlopov [10,11] for the
determination of the probability of PBH formation during the dust-like
phase of the expansion is based on the hypothesis that three eigenvalue
of the deformation tensor are extremely closed to each other:
$\gl_1-\gl_2\simeq \gl_1-\gl_3\sim\gs_{in}$.
Thus the parameters $e$ and $p$ are going to zero,
 and the distribution of the
density field around the point of the maximum is spherical symmetric
with a very good accuracy.
One can obtain the probability of such realization of the
random Gaussian process
 by integration of  Eq.(17) over $\gl_2$ and $\gl_3$. After
integration we have $\gb(M_{bh})\propto\gs_{in}^5$ [10,11].
An additional factor $\gs_{in}^{3/2}$ in Eq.(15) is connected with the
condition of the homogeneous collapse of the configuration according to
the Tolman's solution [10,11].
Taking into account these conditions we would like to
introduce two additional criteria which are important for the
determination of the probability of PBH formation during the dust-like
phase.

First, the collapse of a configuration into a black hole is
definitely a non-local process and a criterium of the field distribution
$\gd$ around a point of a maximum should be non-local also.
Usually, this non-locality is possible to imitate using the filtering
procedure:
\be
\gs(\vec{r},R)=\frac{1}{(2\pi R^2)^{3/2}}\int d^3r' \gd(\vec{r}')
\exp\left[-\frac{(\vec{r}-\vec{r}')^2}{2R_i^2}\right],
\ee
where $R_i$ are different scales of filtering. Let us focus our attention
on the spectrum of perturbations in the PBHs (or another ``dust'' particles,
as was investigated in [10,11]), using Eq.(12).
There exists a low-massive plateau at $M\le M_{eq}$,
 where $\gs_{rms}(M<M_{eq})$
is practically constant for $\ga\ll 1$.

In this interval we use the small scale of filtering $R_s$.
At the same time, at the scale $M\ge M_{eq}$ in terms of Eq.(12) we will
use the large scale filtering $R_l$. Following [10,11], we assume that after
filtering ($R=R_l$) the condition of the spheric symmetry is valid
in the vicinity of some point of a maximum of field $\gd(\vec{r},R_l)$.
The criterium of the homogeneity of distribution $\gd(\vec{r},R_l)$
is  definitely correct for the $R_l$-filtering only.

At the small scale $R\simeq R_s$ the situation is more complicated. Following
[12], it is easy to show that the concentration of the peaks which
is obtained by the
filtering $R\sim R_s$ is much higher than the the concentration of peaks
(not especially high) in $\gd(\vec{r}, R_l)$-field. The correlation function
of the ``smooth scale filtered''  peaks has a typical correlation radius
$r\sim R_s$. If $r\gg R_s$, then the character of the
space distribution of the small
scale peaks is Poissonian. It means that inside a large filtered peak
which is spherical symmetric and homogeneous in average, there are $N\gg1$
small peaks which have a Maxwellian velocity distribution [12].
This configuration (a large scale smoothed peak and a lot of small scale peaks
inside it) has properties to form not a black hole but a virialized cluster.
However, the formation of a cluster  is the natural part of the hierarchical
clustering scenario without any specific and low-probability conditions of
symmetry and homogeneity of the initial configuration.

In the CDM model these clusters are formed as a result of the interaction
of pancakes (see [9]). It is the first reason why the probability
$\gb(M)$ from Eq.(15) overestimates the number of PBHs which can be formed
during the dust-like phase of the expansion of the Universe.

\begin{figure}[h]
\vspace{0.3cm}\hspace{-1cm}\epsfxsize=9cm
\epsfbox{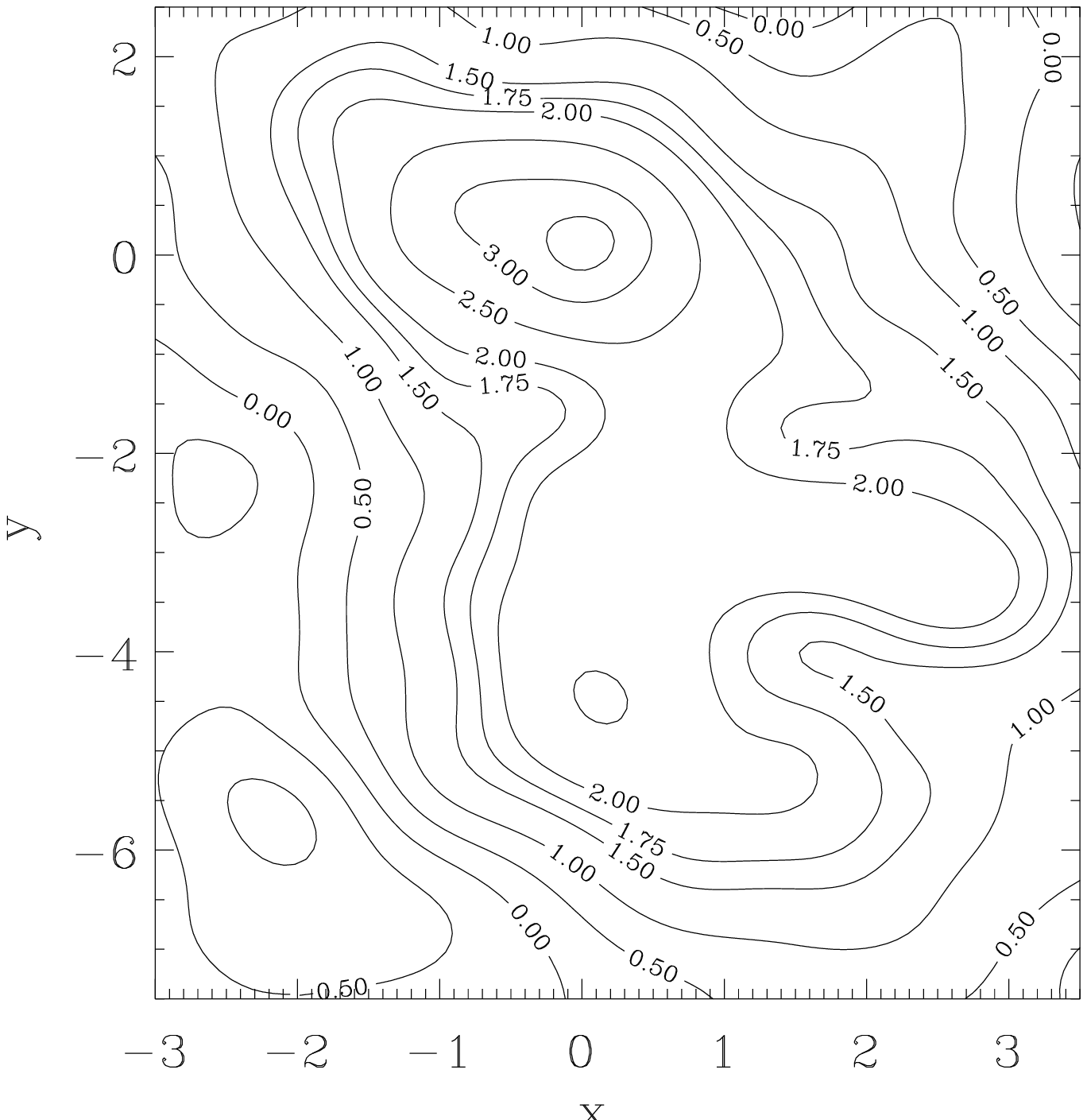}
    \caption{The Distribution of the density field in the vicinity of the
    point of maximum for the typical realization of the two-dimensional
    Gaussian field with Harrison-Zeldovich spectrum of perturbations.
    The origin of the system of coordinates corresponds to the
    point of maximum of $3\gs_{rms}$-peak. The lines represent isodensity
    levels, $x$ and $y$ coordinates are normalized to the filtering radius
    $R_l$.
    There is  a cluster of three peaks: $3.3\gs_{rms}$, $2.6\gs_{rms}$ and
    $2.2\gs_{rms}$. Each member of the cluster determines the curvature and
    symmetry of the isodensity lines in their vicinities at the slice above
    $\ga=2\div2.5$ (see text).
}
\end{figure}

The second reason is the following. The field of peaks which is smoothed
by $R_l$ filter has a peak-peak correlation on a typical scale $r\ge R_l$.
It means that a local criterium of spheric symmetry, which does not
take into account the influence of neighboring peaks, has a limited range
of application in the especially small vicinity of a point of a maximum.
We demonstrate this situation in Fig.2. As it is seen from this figure,
the influence of the neighboring peaks can practically destroy the spherical
symmetry of an initial peak. This problem of an ``outside interaction'' is the
second reason to decreasing of $\gb(M)$-probability.

\section{Induced (coagulation) mechanism of the PBH formation}

In our model the dust-like phase is linked  with the first
generation of the PBHs, and we can introduce the following new mechanism of the
PBH's
formation, based on the typical realization of overdense configurations
 - like pancakes. We don't need a special spheric symmetry and homogeneity
of the configuration because the formation of clusters is the result of
interactions between pancakes similar to the cluster formation
 in the standard CDM model.

Using Press-Schechter formalism we can estimate the moment of formation
of the PBH clusters for the spectrum (11)
\be
t(M_{eq})\simeq
t_{eq}\left(\frac{\gs_{cr}}{\gs_{rms}(M_{eq})T(M_{eq})}\right)^{3/2}
\ll\tau_{ev}(M_{in}),
\ee
where $\gs_{cr}=1.69$. The majority of these clusters are practically
virialized at $t\gg t(M_{eq})$. Let us estimate the characteristic time scale
when the PBHs inside a cluster can collide. The cross-section of the
interaction of PBHs is $\gs_{pbh}=\pi r^2_g$, where $r_g=2GM_{in}$ is the
Schwarzschild radius, and the variance of the velocity is
$v^2_{vir}\sim GM_{eq}/\ov{R}_{eq}$, where $\ov{R}_{eq}$ is the radius of a
cluster.
In this case the time scale of the ``hard core'' collision is:
\be
\tau_{col}\simeq (\gs_{pbh}n_{pbh}v_{vir})^{-1}=
t_o\left(\frac{M_{eq}}{M_{in}}\right)^2
\left(\frac{\gs_{cr}}{\gs_{rms}(M_{in})}\right)^{7/2}.
\ee
Note, that the $\tau_{col}$ is much greater than the time of the velocity
relaxation $\tau_v$ inside a cluster:
\be
\tau_v=\tau_{col}v_{vir}^4\ln ^{-1}\left(\frac{M_{eq}}{M_{in}}\right)\ll
\tau_{col}.
\ee
It means that at  $t\ge \tau_{col}$ the interaction between black holes of
mass $M=M_{in}$ inside a cluster can be described using the coagulation
approximation [14-16].
Let $n(m,t)dm$ be the number of the density of black holes in the mass range
$m\div m+dm$ at the moment $t$. For mass distribution function $n(m,t)$
we can write the coagulation equation [14,15]:
\be
\frac{\pa n}{\pa t}=\frac{1}{2}\int _0^m\ga_c (m',m-m',t)n(m',t)n(m-m',t)dm'-
\ee
\[
-n(m,t)\int_0^{\infty}\ga_c(m,m',t)n(m',t)dm';
\]
where $\ga_c=\lan\gs_{pbh}(m,m')v\ran$, and  $\lan \ \ran$ denotes the
averaging over the peculiar velocity $v$.
After averaging over the peculiar velocity we have
$\ga=\gs_{pbh}(m,m')v_{vir}$.
From  Eq.(25) we can obtain the following expression:
\[
\frac{dM_{tot}}{dt}=\int_0^{\infty} dm'\cdot m'\frac{\pa n}{\pa t}=0;
\]
\be
\frac{dN}{dt}=\int_0^{\infty}dm'\frac{\pa n}{\pa t} =
\ee
\[
=-\frac{1}{2}\int_0^{\infty}dm' \int_0^{\infty}dm\ga_c(m,m')n(m.t)n(m',t),
\]
where $M_{tot}$ is the total mass inside a cluster and $N$ is the total
number of the structured elements in the cluster.
We would like to emphasize that the rate of reaction  $\ga_c$ does not
depend on the time $t$. This approximation is practically the same that was
discussed by Trubnikov [14], and Silk and White [15].
In this case we can use the scaling invariance of the function $\ga_c(m,m')$:
\be
\ga_c(km,km')=k^{\nu}\ga_c(m,m');\hspace{1cm}\nu=2,
\ee
and describe the distribution function $n(m,t)$ as:
\be
n(m,t)=m^{-2}_*(t)F\left(-\frac{m}{m_*(t)}\right)
\ee
where $m_*(t)$ is the typical scale of mass during the process of coagulation
of the PBHs inside the  cluster, and $F$ is some function which depends on
$\left(-\frac{m}{m_*(t)}\right)$. This new function has the following property:
$F\left(-\frac{m}{m_*}\right)\ra 0$ for $m/m_*\ra\infty$.
From Eq.(25)-(27) for $\nu=2$ one can find the equations for $m_*(t)$
and $N(t)$:
\be
\frac{dN}{dt}=-C;\hspace{1cm} \frac{dm_*}{dt}=Dm_*^2,
\ee
where $C$ and $D$ are the constants of normalization.

As it is seen from (29) $N$ decreases with time as:
\be
N=N_0-Ct,
\ee
where $N_0$ is the initial number of points at $t=0$.  In our case
$N_0=M_{eq}/M_{in}$.  For the determination of  constant $C$ we can use
the initial condition of the coagulation process:
\be
n(m,t=0)=N_0\gd(m-M_{in}),
\ee
where $\gd$ is the Dirac $\gd$-function.
Substituting it in Eq.(32) in Eq.(26) we have
\be
\frac{dN}{dt}=-\frac{1}{2}\ga_c(M_{in})N_0^2=-C.
\ee
Let us come back to Eq.(29). As one can find from this equation,
mass of  black holes of the second generation which are formed
inside a cluster increases with time as
\be
M_*=\frac{M_{in}}{1-M_{in}Dt},
\ee
and formally going to infinity at $t=(M_{in}D)^{-1}$. Actually at
$t=\frac{N_0-1}{C}$ the whole mass of a cluster is concentrated in one massive
black hole. But in this limit the coagulation approximation with virialized
velocity field is inapplicable. However, the rate of the coagulation can be
used for qualitative estimation of the mass of a central black hole inside
a cluster. At $t\simeq (N_0-1)/C$ it is equal:
\be
M_{BH}\simeq \frac{M_{in}}{1-\frac{M_{in}DN_0}{C}}=\mu\cdot M_{eq};
\hspace{1cm}\mu\le 1,
\ee
and
\be
D\simeq(2M_{in}\tau_{col})^{-1}\left(1-\frac{M_{in}}{M_{eq}}\right).
\ee

We would like to emphasize that the rate of coagulation (31)-(32) is only
the low limit of the clusterisation of the first generation of the PBHs into
massive primordial black holes
 of the second generation. It is caused by the taking into account
only ``hard-core'' collisions of black holes inside a cluster.
In a general case  the formation of pairs is important and the rate of
coagulation increases\footnote{We describe this situation in a separate
paper.}.
Thus if $\tau_{col}\ll\tau_{ev}(M_{in})$, then the PBHs form
a massive black hole with $M\gg M_{in}$ inside a cluster.
In Fig.1 we plotted the thick solid
line with diamonds which corresponds to the equality
$\tau_{col}=\tau_{ev}(M_{in})$.
If initial amplitudes of perturbations at the range $M_{in}<10^{14}$g are
$\gs_{rms}(M_{in})\ge 0.1$ then the first generation of the PBHs produces
 the dust-like
phase and induce the formation of the second generation of black holes.
Two thick lines with asterisks correspond to the condition $M_{eq}<10^9$g
(left line) and $M_{eq}<10^{15}$g (right line). In the first case $M_{in}$
has to be less than $10^5M_{Pl}$. At the second case we have
$M_{in}<10^7M_{Pl}$. In both cases the critical level of the power index
$n$ is $n=1.23$. What will happen if this condition is broken?
Obviously the first generation of PBHs forms the second generation of
PBHs  with $10^9$g$<M<10^{15}$g and $\gb(10^9$g$<M<10^{15}$g$)\le 1$.
We can estimate this factor $\gb$
using the Press-Schechter formula. For the clusters with $M\approx M_{eq}$ the
concentration is the following
\[
n(M,t)dM=
\frac{4}{3\sqrt{\pi}}\frac{\ov{\gr}_{pbh}}{M^2}
\left(\frac{M}{M_{max}(t)}\right)^{2(1+\ga)/3}\times
\]
\be
\times\exp\left[-\left(\frac{M}{M_{max}(t)}\right)^{2(1+\ga)/3}\right]dM,
\ee
where $M_{max}$ is determined by Eq.(14), and $\ov{\gr}$ is the mean
density of the PBHs during the dust-like phase. This density of the
clusters with masses $M_{eq}\le M\le M_{max}$ is
$\gr_{cl}\sim \ov{\gr}_{pbh}$.
If each cluster of mass $M_{eq}$  contains a massive black hole with
the mass $M_{bh}=\mu M_{eq}$, $\mu\le 1$, then the probability of the
second generation black hole formation is
\be
\gb_{pbh}^{II}(M_{bh})\simeq \mu \gg \gb_{pbh}^{obs}
\ee
where $\gb_{pbh}^{obs}$ is the observational limit\footnote{We would like
to note that $\mu$ is remarkable less than unity, for example due to emission
of the gravitational waves [17]}.

\section{Conclusions}
We have presented here the new mechanism of the formation of the primordial
black holes in the very early Universe after inflation.
We have shown that a high level of adiabatic
perturbations in the early Universe leads to producing the first generation
of the PBHs. Further these PBHs form the specific phase of the cosmological
expansion - so called dust-like phase.
The hierarchical clustering of perturbations
in a ``gas'' of the PBHs is the source of formation of the next generation
of the black holes inside the clusters. We have presented the new mechanism
of the
second generation of the PBHs formation which is based on the coagulation
mechanism. We have shown that the initial inhomogeneity of the Universe
``switch on'' the mechanism of the induced PBH formation and leads to their
over-producing if the power index of perturbations is $n\ge 1.2$.
We note that the formation of PBH from the ''point-like'' collisionless
particles  and from PBH of first generation has only one difference.
Using Eq.(24) one can see, that $\tau _{col}\propto M^{-2}$ and
goes to infinity, if $M=m\rightarrow 0$ , where m is the mass of
particles. In this case the coagulation mechanism of the PBH
formation is inapplicable and the probability of the PBH formation
is extremely low.

{\large\bf Acknowledgments}

The authors are grateful to A.Doroshkevich and I.Novikov for useful
discussions.
 P.N. is grateful to the staff of TAC, for providing
excellent working conditions. This investigation was supported
by the Danish Natural Science Research Council
through grant No. 9401635 and also in
part by Danmarks Grundforskningsfond through its support for the
establishment of the Theoretical Astrophysics Center.
\newpage
{\large\bf References}

\begin{enumerate}
\item\
S.Mollerach, S.Matarrese, F.Lucchin, Phys. Rev. D. {\bf 50}, 4835, (1994);
J.Garcia-Bellido, A.Linde and D. Wands, Phys. Rev. D. {\bf 54}, 6040, (1996);
J.Garcia-Bellido, A.Linde, Phys. Lett. B {\bf 398}, 18, (1997);
J.Garcia-Bellido, A.Linde, Phys. Rev. D. {\bf 55}, 7480, (1997).
\item\
C.L.Bennett et al. Astrophys. J. Lett. {\bf 464} L1 (1996).
\item\
B.J.Carr, J.H.Gilbert, J.E.Lidsey, Phys. Rev. D. {\bf 50}, 4853, (1994).
\item\
A.M.Green and A.Liddle, Astro-ph. 9704251.
\item\
B.J.Carr, Astrophys. J. {\bf 205}, 1 (1975);
Ya.B.Zeldovich, A.A.Starobinsky, M.Yu.Khlopov and V.M. Chechetkin,
       Pis'ma Astron. Zh. {\bf 3}, 308, (1977) [Sov. Astron. Lett {\bf 22},
       110 (1977)].
S.Mujana and K.Sato, Prog. Theor. Phys. {\bf 59}, 1012 (1978);
B.V.Vainer and P.D.Naselsky, Astron. Zh, {\bf 55}, 231 (1978) [Sov.
       Astron. {\bf 22}, 138 (1978)];
B.V.Vainer, O.V.Dryzhakova and P.D.Naselsky, Pis'ma Astron. Zh. {\bf 4}, 344
       (1978)[Sov. Astron. Lett. {\bf 4}, 185 (1978)];
I.D.Novikov, A.G.Polnarev, A.A.Starobinsky and Ya.B. Zeldovich, Astron.
	Astrophys. {\bf 80}, 104 (1979);
D.Lindley, Mon. Not. R. Astron. Soc. {\bf 193}, 593 (1980);
T.Pothman and R. Matzner, Astrophys. Space. Sci. {\bf 75}, 229 (1981);
J.H.MacGibbon, Nature {\bf 320}, 308 (1987);
J.H.MacGibbon, and B.Carr, Astrophys. J. {\bf 371}, 447 (1991).
\item\
J.R.Bond. Phys.Rev.Lett.,  {\bf 74}, 4369, 1995.
\item\
B.J.Carr and S.Hawking, Mon. Not. R. Astron. Soc. {\bf 168}, 399 (1974).
\item\
E.F.Bunn, A.R.Liddle and M.White, Phys. Rev. D {\bf 54}, 5917, (1996).
\item\
A.G.Doroshkevich and E.V.Kotok, Mon. Not. R. Astron. Soc. {\bf 246}, 10,
(1990);
A.G.Doroshkevich, R.Fong,  S. Gottl\"{o} ber and J. M\"{u} kket,
 Mon. Not. R. Astron. Soc. {\bf 284}, 663 (1996);
A.G.Doroshkevich, et al., Mon. Not. R. Astron. Soc. {\bf 284}, 1281
        (1996);
M.Demianski and A.G.Doroshkevich Astrophys. J. submitted (1997).
\item\
A.G.Polnarev and M.Yu.Khlopov, Phys. Lett. B {\bf 97}, 383, 1980.
\item\
A.G.Polnarev and M.Yu.Khlopov, Sov. Astron, {\bf 26}, 391, 1983.
\item\
J.M Bardeen, J.R.Bond, N.Kaiser and A.S.Szalay, Astrophys. J.,
    {\bf 304}, 15, (1986)
\item\
A.G.Doroshkevich, Astrophysica {\bf 6}, 320, (1970).
\item\
B.A.Trubnikov, Soviet Phys. Doklady {\bf 16}, 124, (1971).
\item\
J.Silk and S.D.White, Astrophys. J., {\bf 223}, L59, (1978).
\item\
J.Silk and C.Norman, Astrophys. J., {\bf 247}, 59, (1981).
\item\
Ya.B.Zeldovich and I.D.Novikov, Doklady Akad. Nauk {\bf 155}, 1033 (1964).
see also J.Baker et al. gr-qc/9608064.
\end{enumerate}
\end{document}